\documentclass[aps,prd,twocolumn,superscriptaddress]{revtex4}
\usepackage{graphicx,amsmath}
\usepackage[latin1]{inputenc}
\usepackage[english]{babel}
\usepackage[T1]{fontenc}
\usepackage{amsmath,amssymb}

\graphicspath{{./figs/}}
\usepackage{sistyle}
\usepackage{wasysym}
\usepackage[final,danish]{fixme}
\usepackage{hyperref} 

\renewcommand{\d}{\text{d}}
\newcommand{\fref}[1]{Fig. \ref{#1}}
\newcommand{\eref}[1]{Eq. ~(\ref{#1})}
\newcommand{\tref}[1]{Tab. \ref{#1}}

\begin{document}

\title{Experimental investigations of synchrotron radiation at the onset of the quantum regime}
\author{K.K. Andersen, J. Esberg, H. Knudsen, H.D. Thomsen, U.I. Uggerh{\o}j}
\affiliation{Department of Physics and Astronomy, University of Aarhus, Denmark}
\author{P. Sona}
\affiliation{Department of Physics and Astronomi, University of Florence, Florence, Italy}
\author{A. Mangiarotti}
\affiliation{Universidade de Coimbra, Portugal}
\author{T.J. Ketel}
\affiliation{VU University, Amsterdam, The Netherlands}
\author{A. Dizdar}
\affiliation{University of Istanbul, Istanbul, Turkey}
\author{S. Ballestrero}
\affiliation{University of Johannesburg, Johannesburg, South Africa}



\collaboration{CERN NA63}

\date{\today}

\begin{abstract}
The classical description of synchrotron radiation fails at large Lorentz factors, $\gamma$, for relativistic electrons crossing strong transverse magnetic fields $B$. In the rest frame of the electron this field is comparable to the so-called critical field $B_0 = 4.414\cdot10^9$ T. For $\chi = \gamma B/B_0 \simeq 1$ quantum corrections are essential for the description of synchrotron radiation to conserve energy.
With electrons of energies 10-150 GeV penetrating a germanium single crystal along the $\langle110\rangle$ axis, we have experimentally investigated the transition from the regime where classical synchrotron radiation is an adequate description, to the regime where the emission drastically changes character; not only in magnitude, but also in spectral shape. The spectrum can only be described by quantum synchrotron radiation formulas. Apart from being a test of strong-field quantum electrodynamics, the experimental results are also relevant for the design of future linear colliders where beamstrahlung - a closely related process - may limit the achievable luminosity. 
\end{abstract}


\maketitle



\section{Introduction}
In the seminal paper by Schwinger on emission of synchrotron radiation \cite{Schw49}, the classical treatment is shown to be invalid once the momentum of the emitted quantum becomes comparable to the electron momentum, i.e.\ the condition for validity is $\hbar\omega\ll E,~\omega\sim\omega_c$ where $E=\gamma mc^2$ is the energy of the electron, $\gamma$ the Lorentz factor of the electron, $\hbar\omega$ the energy of the photon and $\omega_c$ the critical frequency of the emitted synchrotron radiation. This condition - not included in his original manuscript \cite{Schw45} - he wrote as a function of the magnetic field $B$ as $E/mc^2\ll mc^2/((e\hbar/mc)B)$, which translates into $\chi\equiv\gamma B/B_0\ll 1$ where $B_0$ is the critical magnetic field, $B_0=m^2c^3/e\hbar=4.414\cdot10^9$ T. The equivalent electric field is $\mathcal{E}_0=m^2c^3/e\hbar=1.32\cdot10^{16}$ V/cm. As also remarked by Schwinger, with a normally obtainable field of up to 1 T, classical theory is adequate up to energies of $10^{15}$ eV, even nowadays an extreme energy. However, since $\chi$ is a relativistic invariant, the combination of strong fields and high Lorentz factors may give rise to synchrotron radiation in the quantum regime where $\chi\gtrsim1$.

In crystals, electric fields of the order $10^{11}$ V/cm, corresponding to magnetic fields of a few $10^4$ T and originating from the screened nuclear fields, are achievable along crystallographic axes (see below). Thus, using particles with $\gamma\sim10^5$ as e.g.\ the 100 GeV electrons obtainable at the CERN SPS, experimental investigations of synchrotron radiation in the quantum regime become possible.


\subsection{Spin-flip transitions}
Not only does the total radiated energy differ substantially between the classical and quantum regimes, the spectral composition becomes drastically different once the energy of the impinging particle gets sufficiently high. Part of the explanation for this comes from the increasing relevance of spin-flip processes in the quantum regime. Spin-flip processes are slow and of low energy for $\chi\ll1$, but become fast and extremely energetic for $\chi$ comparable to or beyond 1. In a simple classical picture, one can explain why the energy of a spin-flip transition becomes comparable to the electron energy for $\chi\sim1$. 
Due to the Lorentz transformation to the rest frame of the penetrating electron, the strong electric field in the crystal appears as a strong magnetic field $B=\gamma\beta {\mathcal E}_{\mathrm{lab}}$, in which 
the magnetic moment of the electron achieves an energy of $E_{\mu} = -\vec{\mu}\cdot\vec{B}$, giving rise to emission connected to a spin-flip transition. In an 'elementary treatment', which is not completely correct but 'has direct intuitive appeal and works surprisingly well for electrons' \cite{Jack76}, the spin-flip transitions of electrons with $\mu=e\hbar/2mc$ have an energy $\Delta E_{\mu}=e\hbar B/mc$ in the rest system \cite{Kirs01a}. Transformation back to the laboratory yields a factor $\gamma$ such that the result is
\begin{equation}
\Delta E_{\mu}=\gamma^2\beta\frac{{\mathcal E}}{{\mathcal E}_0}mc^2=\chi\beta\gamma mc^2
\end{equation}
so the electron radiates a significant fraction of its energy, $E$, when $\chi\beta\simeq1$. This simple estimate shows why spin-flip radiation transfers large fractions of the kinetic energy when $\chi\sim1$ is reached. However, the model has obvious limitations since it predicts energy transfers larger than the electron energy for $\chi \simeq 1$.

The time scale of spin-flip transitions, $\tau_\text{sf}$, that lead to polarization of the electron is given by \cite{Soko86,Baie70,Schw74}
\begin{equation}
\tau_\text{sf}=\frac{8\hbar}{5\sqrt{3}\alpha mc^2}\frac{\gamma}{\chi^3}
\label{pol_time}
\end{equation}
which may be rewritten as $c\tau_\text{sf}=\varepsilon_\text{sf}\gamma a_0/\chi^3$ where $\varepsilon_\text{sf}=8/5\sqrt{3}\simeq92.4\%$ is the maximum polarization degree due to spin-flip transitions and $a_0$ is the Bohr radius.
For a 100 GeV electron in a $\chi=1$ field, $c\tau_\text{sf}$ becomes 10 $\mu$m or $\tau_\text{sf}=32$ fs. For the more usual situation of a 1 GeV electron in a 1 T field, $c\tau_\text{sf}$ is 7.3 astronomical units and $\tau_\text{sf}$ is 61 minutes - a typical polarization time in an accelerator.
Therefore a substantial fraction of the radiation events originate from spin-flip transitions as one gets to and beyond $\chi\simeq1$, even in a target as thin as 0.1 mm. Theoretical studies \cite{Bary89} point to the possibility of obtaining polarized positrons in a similar manner, by planar channeling through bent crystals of lengths about 1 mm.

Ultrarelativistic leptons with $\gamma\gtrsim10^5$ that penetrate a crystal where they are exposed to fields $\mathcal E\simeq10^{11}$ V/cm may therefore probe the quantum regime $\chi\gtrsim1$ where spin-flip transitions and recoil becomes decisive for the intensity and shape of the radiation spectrum. See e.g.\ \cite{Baie98,Ugge05b} for reviews on such strong-field effects in crystals.


\subsection{Beamstrahlung}
Synchrotron radiation in the quantum regime is of great importance to the design of the collision point and a potentially serious limitation for the next generation linear colliders based on lepton beams. The crucial phenomenon here is the emission of intense radiation due to the interaction of one bunch with the electromagnetic field from the opposing bunch. The particle deflection imposed by the field of the oncoming bunch leads to an emission process very similar to synchrotron radiation: beamstrahlung. As the emission of beamstrahlung has a direct and significant impact on the energy of the colliding particles, it is a decisive factor for e.g.\ the energy-weighted luminosity. Conversely, beamstrahlung emission may provide a method for luminosity measurement. It is therefore important to know if beamstrahlung theory is correct, for the conceptual and technical design of the collision region - the interaction center about which the rest of the machine is based.

The Lorentz factor $\gamma$ in the case of a collider is understood as the Lorentz factor of each of the oppositely directed beams, measured in the laboratory system. Then relativistic velocity addition yields the Lorentz factor $\gamma'$ - responsible for length contraction or time dilation - of one beam seen from a particle in the other beam of $\gamma'=2\gamma^2-1$, usually shortened to $2\gamma^2$ in the ultrarelativistic limit. Thus, in the rest frame of one bunch the field of the other bunch is boosted by a factor $\simeq2\gamma^2$ and may approach or even exceed critical field values.

The emission of beamstrahlung can be expressed as a function of $\chi$ (often called $\Upsilon$ in the accelerator physics community) which is of the order of  unity for the next generation linear colliders \cite{Chen88}. For the planned Compact LInear Collider (CLIC) at CERN, the collision point is designed such that the average value of the strong-field parameter is $\overline{\Upsilon}\simeq4$ and for the International Linear Collider (ILC) with 500 GeV center-of-mass energy $\overline{\Upsilon} = 0.045$. For CLIC, due to the emission of beamstrahlung, the peak luminosity $\mathcal{L}_1$ (where $\mathcal{L}_1$ is defined as the luminosity for that part of the beam where the energy is still at least 99\% of the initial value) becomes only about 34\% of the nominal. Simulations show that if the radiation probability was not suppressed by quantum effects, the peak luminosity $\mathcal{L}_1$ would be further reduced to 17\% of the nominal value. This illustrates the importance of the inclusion and understanding of strong field effects in the design of a future collider. For these future colliders, $\gamma\gamma$-collisions, resulting in e.g.\ hadronic interactions, may be generated from the beams themselves and the advantage of using leptonic beams, that give 'clean' collisions, is abated. The beamstrahlung problem is unavoidable since small beam cross sections are needed to give high luminosity for single passage machines (as opposed to circular machines). Since $\Upsilon\propto N\gamma\cdot\sigma_z/(\sigma_x+\sigma_y)$, with $N$ the number of particles in each bunch and $\sigma$ denoting the beam size, with high energies and high luminosity a high value of $\Upsilon$ is obtained. However, the problem may be partly alleviated by applying special bunch structures (ribbon pulses) to avoid rapid beam deterioration from strong-field effects \cite{Blan88b}.

Thus, design schemes for the next generation colliders rely heavily on calculations of synchrotron radiation in the transition region from the classical to the quantum regime, $\chi\simeq1$. However, apart from a previous study performed with a tungsten target without the ability to measure the spectral composition of the emitted photons \cite{Kirs01a}, such calculations have hitherto not been experimentally tested in the relevant region of values of $\chi$.\\

In the following we show experimentally, that when exposed to the strong, pseudo-continuous field in a crystal, the radiation changes character, with the intensity going from the classical $\gamma^2$ dependence towards a much weaker dependency, consistent with a theory predicting that eventually it becomes proportional to $\gamma^{2/3}$. The field, in this case, is 'turned on' when the crystal is rotated from a 'random' orientation where the scattering centers act incoherently to an axial orientation where the fields are coherently adding up along the direction of motion of the penetrating particle.


\section{Theoretical expectations}
Here we will briefly introduce some theoretical aspects and definitions that are neccessary to get an understanding of the radiation process in the high-field environment of a crystal axis. The discussion will only concern axial strong fields and all explicit values are for the germanium $\langle 110 \rangle$ axis at room temperature. These can be found in \cite{Baie98}.

Let $\theta$ be the angle between the direction of an electron and the crystal axis. Radiation can be divided into several regimes depending on $\theta$. Electrons that have entrance angles smaller than the Lindhard critical angle $\theta_c$ are trapped in the cylindrical potential along the crystal axis or are in low above-barrier states. This angle is given by $\theta_c = \sqrt{2V_0/E}$ which is $\SI{47}{\mu rad}$ at $E = \SI{100}{GeV}$, where $V_0 = 110$ eV is the transversal  potential height of the crystal. Particles with large entrance angles will on the other hand have a transversal kinetic energy that is higher than the crystal potential and will approximately follow rectilinear paths. The parameter $\rho(\theta_0) \simeq (2V_0/mc^2\theta_0)^2$ \cite{Baie98} compares the particle deflection angle to the characteristic radiation angle. For large $\rho$ the particle deflection is larger than the radiation cone and the particle satisfies the so-called magnetic bremsstrahlung (synchrotron radiation) condition. Oppositely, for small deflections the radiation spectrum satisfies the dipole approximation. The characteristic angle dividing these two regimes is independent of the electron energy and given by $\theta_v = V_0/mc^2 = \SI{215}{\mu rad}$ which is called the Baier angle. 

\begin{figure}[tb]
\centering
	\includegraphics[width=1\columnwidth]{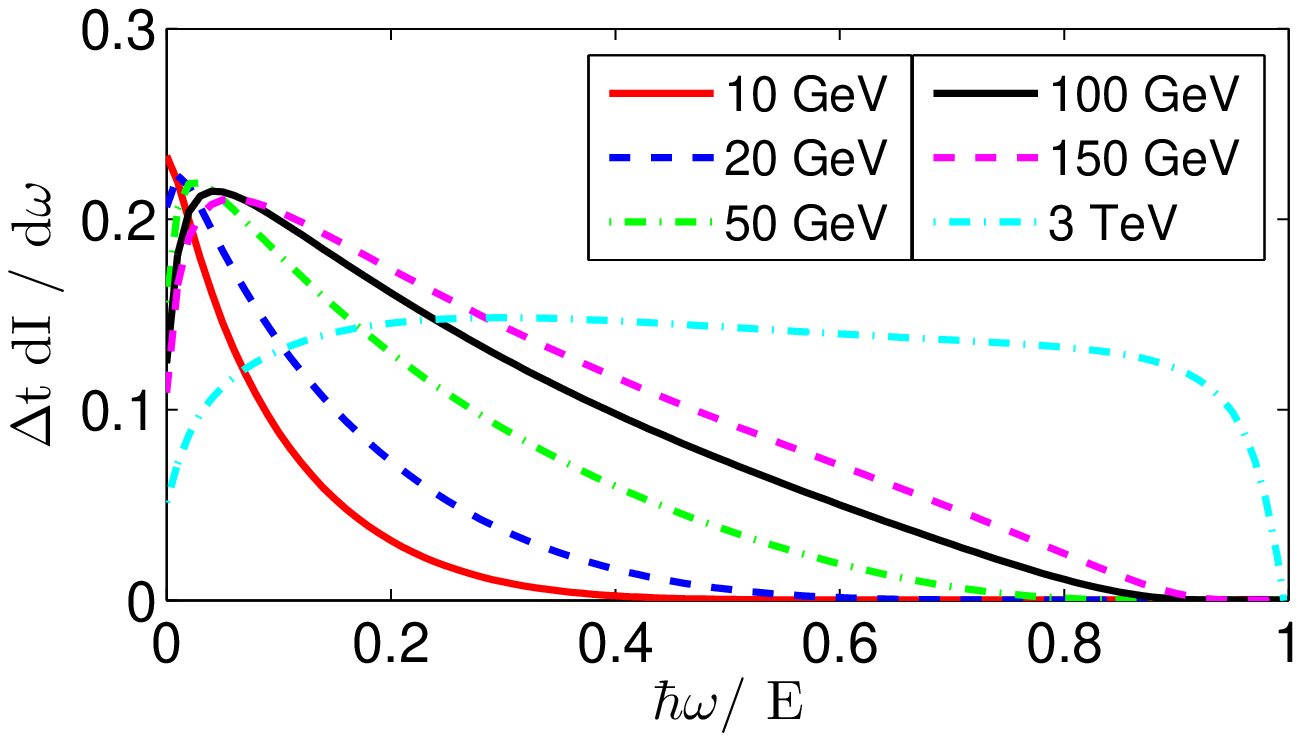}
	\caption{\label{fig:dI} The differential radiation spectrum from an electron traversing a Ge $\langle110\rangle$ crystal at 280 K entering the crystal along the axis for different electron energies. The curves have been calculated with Eq. (17.7) in \cite{Baie98}. $\Delta t= \Delta l/c$ is the time it takes to traverse the $\SI{200}{\mu m}$ crystal used in the experiment.}
\end{figure}

\begin{figure}[tb]
\centering
	\includegraphics[width=1\columnwidth]{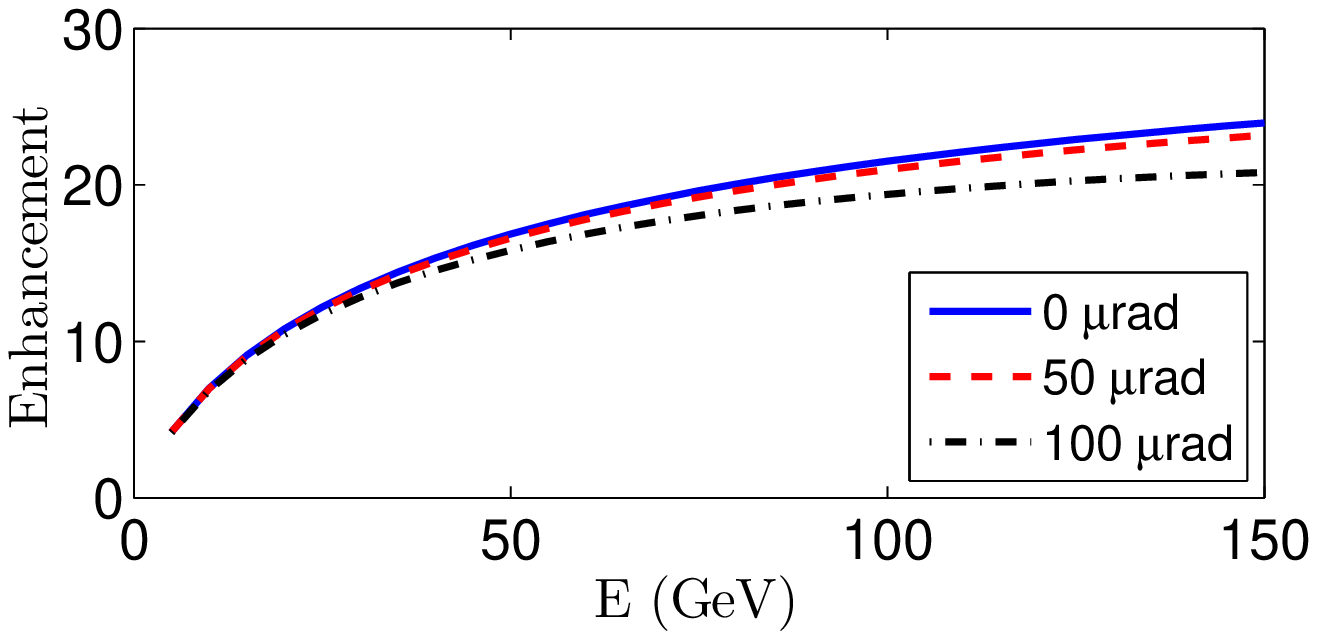}
	\caption{\label{fig:enhteo} The integral radiation intensity in axial alignment (calculated with Eq. (17.15) \cite{Baie98}) divided by the Bethe-Heitler radiation for different electron entrance angles and as a function of the energy $E$ of the impinging electron.}
\end{figure}

In the present experiment we investigate radiation in the magnetic bremsstrahlung limit where one can essentially regard all radiation as emitted from a constant field. This regime is obtained for $\theta \ll \theta_v$ and $\theta_\gamma \ll \theta_v$ where $\theta_\gamma = 1/\gamma$ is the radiation angle which is e.g. $\SI{5}{\mu rad}$ for a 100 GeV electron. For $\theta \leq \theta_c$ redistribution of the electron flux is important since the electrons are mainly confined to the axis. This affects the radiation yield significantly. For a broad $\theta$ range, the effect of the redistribution is unimportant but the constant field approximation (CFA) is still valid.

Formulas for the differential radiation spectrum, $\d I/\d\omega$, from electrons at our experimental conditions and the total radiation intensity, $I_\text{tot}$, are given by Baier \textit{et al.} \cite[Eq. (17.7)]{Baie98}. The differential radiation spectrum for electrons moving along the axis is shown in \fref{fig:dI} and the integral radiation enhancement as a function of the electron energy and entrance angle in \fref{fig:enhteo}. 

The integral radiation enhancement is defined as 
\begin{equation}
\eta_\text{tot} = \frac{I_\text{tot}}{E}\times X_0,
\label{eq:etatot}
\end{equation}
where $X_0$ is the radiation length of the crystal. For simplicity and since the beam divergence is large ($\sim \SI{100}{\mu rad}$) we neglect the effect of redistribution and assume that no particles are trapped in states around a crystal axis (in other words no particles are channelled). 

The strong-field parameter $\chi$ as defined in the abstract corresponds to a particle exposed to a constant field. However, for particles penetrating a crystal along an axis or plane, the particles experience different fields depending on the distance to the axis. Therefore one cannot probe a single value of the $\chi$ parameter with a crystal but instead measures a combined effect. For crystals it is useful to define the associated strong-field parameter \cite{Baie98} 
\begin{equation}
\chi_s = \frac{V_0E \hbar c}{a_s(mc^2)^3}.
\label{eq:chis}
\end{equation}

\begin{figure}[tb]
	\centering
		\includegraphics[width=1.00\columnwidth]{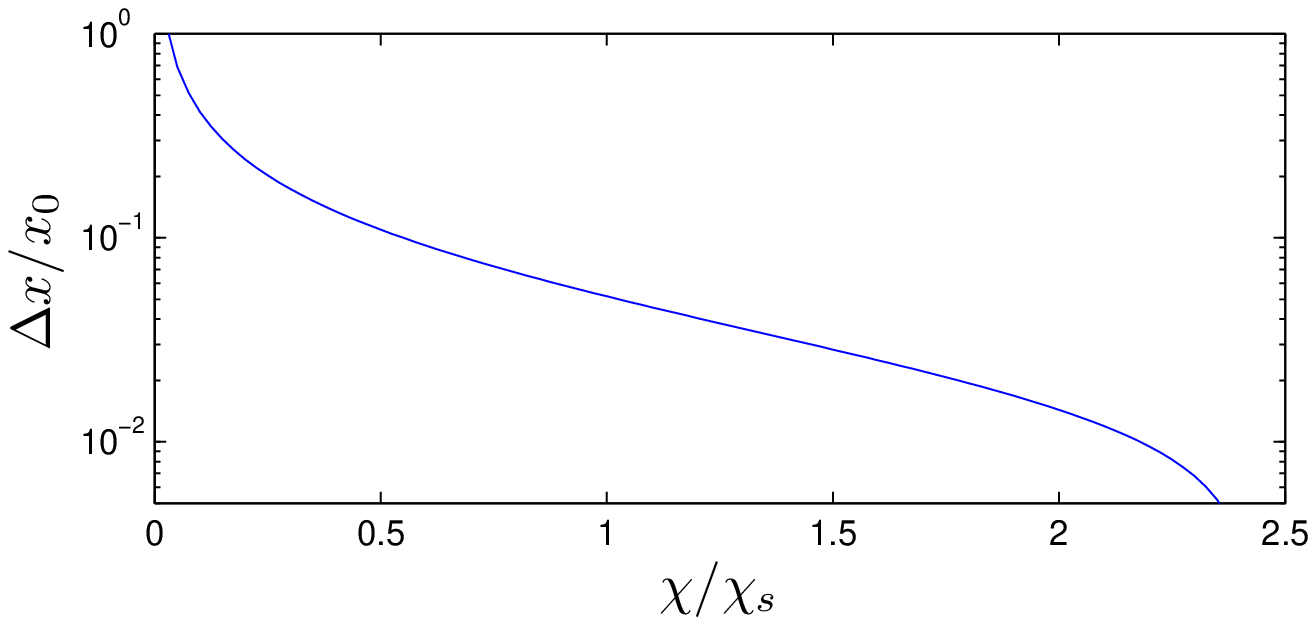}
	\caption{	\label{fig:chifrac} For a uniform particle distribution traversing a Ge crystal along the $\langle 110 \rangle$ axis is shown the fraction of particles that experience a field $\chi/\chi_s$ or above.  }
\end{figure}

The local strong-field parameter as a function of the distance to the axis is \cite[Eq. (17.10)]{Baie98}
\begin{equation}
\chi(x) = 2\chi_s\frac{\sqrt{x}}{(x+\eta)(x+\eta+1)},\  0\leq x \leq x_0, \ x = \rho^2/a_s^2
\label{eq:chi}
\end{equation}
where $\eta = 0.115$, $a_s=\SI{0.337}{\angstrom} $, and $x_0 = 15.8$ are crystal parameters. $a_s$ is the screening distance and $x$ is the square of the distance to the crystal axis in units of $a_s$. It is clear from \eref{eq:chi} that ${\chi}$ is proportional to $\chi_s$. For a uniform particle distribution one can find the fraction of particles that experience a $\chi$-value of $y = \chi/\chi_s$ or above. This is plotted in \fref{fig:chifrac}. It can be seen that only about 10 \% of the particles travel in  $\chi \geq 0.5 \chi_s$.

\section{Experimental setup}
\begin{figure*}[tb]
	\centering
		\includegraphics[width=1.00\textwidth]{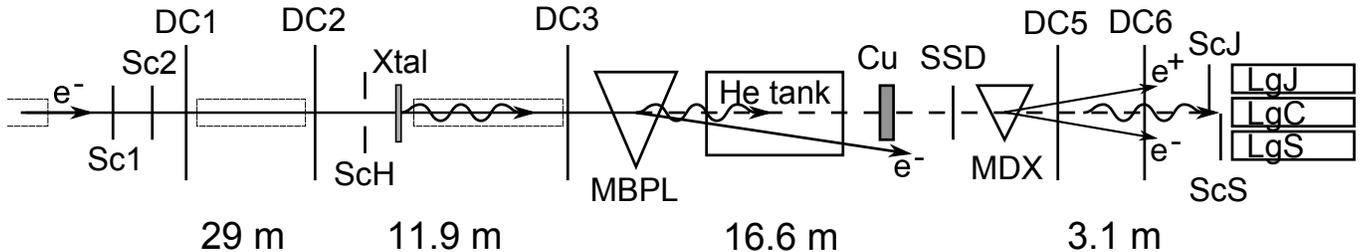}
	\caption{\label{fig:setup} Experimental setup (not to scale). An electron beam enters from the left. DC refers to drift chambers and Sc to scintillators. Xtal is the Ge crystal ($0.87 \% X_0$), ScH a hole scintillator ($\diameter\SI{12}{mm}$), MBPL the deflection magnet, Cu a copper converter, SSD a solid-state detector, MDX the second deflection magnet and Lg refers to lead glass calorimeters (J for Jura and S for Salève, landmarks near Geneva, and C for center). Dashed boxes indicate vacuum pipes. The overall length is around 60 meters from the first scintillator to the lead glass detectors. All distances on the figure are between adjacent DCs.}
\end{figure*}

The experiment was carried out using a tertiary beam of electrons from the CERN SPS at the H4 beamline in the North Area. The energy of the beam was varied from 10 GeV to 150 GeV. The experimental setup is shown in \fref{fig:setup} (the figure is not to scale). The beam is defined by hits in scintillators Sc1 and Sc2 and no hit in ScH which has a $\diameter 12$ mm hole. The position of the beam is determined with drift chambers 1 - 3 (DC1-DC3) with a resolution of approximately 0.17 mm (FWHM). The distance from DC1 to DC2 is $\SI{29}{m}$ giving an entrance angle sensitivity of $\sim \SI{20}{\mu rad}$ which is significantly smaller than the deflection caused by multiple coulomb scattering (MCS) which is of the order $\sim \SI{180}{\mu rad}$. This is mainly caused by two scintillators used for beam control (not shown in \fref{fig:setup}). DC2 and DC3 are used to measure the direction of the electron after passing the $\SI{200}{\mu m}$ Ge crystal Xtal ($0.87\% X_0$). We assume that this is the direction of the emitted photon, correct up to an angle $\sim1/\gamma$. With electron energies from 10 GeV to 150 GeV this corresponds to angles from $\SI{50}{\mu rad}$ to $\SI{3}{\mu rad}$, which is comparable to or less than the detection resolution. 

Downstream of DC3, the electron beam is deflected by a dipole magnet (MBPL). After the deflection, photons and electrons travel 10 m in a helium bag for the beam to be sufficiently separated from the photons.
A copper target is inserted after the helium bag. This is part of the pair spectrometer (PS) and has the purpose of converting photons to electron-positron pairs. The thickness of the converter is 14\%$X_0$ which is a compromise between getting good statistics and avoiding events where two or more photons convert. The pair-conversion probability is constant for the relevant photon energies. Hence, the spectrum measured with the pair spectrometer corresponds to a pile-up free photon spectrum scaled by a conversion factor. A solid-state detector (called SSD, $\diameter \SI{5}{cm}$) is inserted after the Cu converter. It can be used to check if any pairs were created in the converter or further upstream. All pairs created in the converter and with energies above 100 MeV will be detected by the SSD detector since the Borsellino angle \cite{Bors53} (the angle between the electron and the positron of the created pair) is given by $\theta = \frac{4mc^2}{\hbar\omega}$ which is $20$ mrad for a 100 MeV photon and the distance from the converter to the detector is $\sim\SI{1}{m}$. 

A MDX dipole magnet deflects pairs generated between the MBPL and the MDX. The deflection of the pair is measured by DC5 and DC6, from which the momenta of the particles can be found. The PS is described in detail below. Downstream of DC6 are three lead glass calorimeters (LG). These are used to detect the energetic pairs (deflected less than $\SI{26}{mrad}$) and the majority of the photons ($\sim 89\%$) which do not convert and hit the central lead glass calorimeter (LgC).  

With the pair spectrometer we can remove pile-up from the photon spectrum which is not possible with the lead glass calorimeter. Pile-up changes the shape of the true radiation spectrum mainly by decreasing the low-energy part but also by slightly increasing the high-energy part. The disadvantage of the pair spectrometer is lower statistics and a higher detection threshold ($\sim10$ GeV, see below) compared to the lead glass calorimeter.

To calculate the momentum of a pair we demand 2 hits in DC5 and 1 or 2 in DC6 (or the reverse) and hits in 2 of the 3 first DCs. Hits in the DCs before the MBPL are used to define the direction of the emitted photon. In the case of two hits in both DC5 and DC6, we can directly find the angles of the electron and positron and use this to calculate the deflection caused by the MDX.  In the case where the electron only hits one of DC5 and DC6, we extrapolate the trajectory from DC1-3 and the positron hits in DC5 and DC6 to find the deflection vertex in the MDX magnet. The vertex is defined as the position where the extrapolated trajectories from the drift chambers upstream and downstream intersect in the MDX magnet. The deflection of the electron is then found from the coordinates of the DC hit and the deflection vertex. A similar procedure can be done in the case where the positron only hits one DC. Since the detection efficiency of the drift chambers is high ($\sim 95\%$) these events dominate for low energy particles where a particle is detected in DC5 but deflected outside DC6.

The integrated field of the magnet is calibrated by measurements of the deflection of beams of known energies. From the deflection measured by DC5 and DC6 the integrated magnetic field of the MDX magnet was found to be in good agreement with standard magnet calibrations performed by the use of probes. Furthermore, we have estimated the PS momentum resolution as a function of particle momentum. We find $\frac{\delta p}{p} = \sqrt{A^2 + B^2p^2c^2}$ with $A = 0.051$ and $B = \SI{0.57e-3}{GeV^{-1}}$, where $A$ depends on the amount of MCS from DC5 to DC6 and is in good agreement with expectations. The $B$ term is connected to the chamber resolution and reflects a position uncertainty of $\sim\SI{170}{\mu m}$. 
Thus, for converted photons, the measured deflections may be used to calculate the momenta of the electron and positron, and subsequently the energy of the radiated photon.

\subsection{Germanium crystal}
The germanium crystal was mounted on a goniometer for a stable and precise orientation of the crystal. The rotation of the crystal can be controlled in $\SI{1.7}{\mu rad}$ steps. The crystal was aligned by rotations around the vertical and horisontal axis (transverse to the beam direction) and measuring the radiation yield. After the experiment a thorough analysis of the scanning data was performed. The results show that the position defined as ``on axis'' during the experiment was approximately $\SI{0.08}{mrad}$ away from the real axis. We have investigated the effect of the displacement by plotting the enhancement as a function of the entrance angle. Since the strong-field effects are present within the Baier angle we do not observe any significant variation of radiation enhancement for entrance angles less than $\SI{0.2}{mrad}$.

Furthermore, the crystal setup was inadvently perturbed so the position of the axis changed during the experiment. When this was discovered a new axis position was determined. The post analysis shows that the axis position had changed around $\SI{0.3}{mrad}$. The 50 GeV data were measured right before this was discovered, and one cannot exclude that they might be affected. However, $\SI{0.3}{mrad}$ is not significantly larger than the Baier angle $\theta_v$ inside which the strong-field effects are only weakly dependent on the exact direction.

The thickness of the crystal has been measured to be $\SI{180}{\mu m}$ at one edge and $\SI{220}{\mu m}$ at the opposite. From X-ray reflection a clear $\langle 110 \rangle$ axis has been observed with a mosaic spread of $\sigma_\text{mosaic} = \SI{0.22}{mrad}$. This is comparable to the Baier angle. The mosaic spread is a source of dechannelling which will slightly reduce the radiation enhancement. However, this is not important for this experiment, where only a small fraction of the electrons are channelled. Furthermore, channelling leads to redistribution of the particles and this is not included in the theoretical calculations. Hence, the mosaic spread is of negligible importance for these measurements.


\section{Data analysis}

\begin{figure*}[tb]
	\centering
		\includegraphics[width=0.8\textwidth]{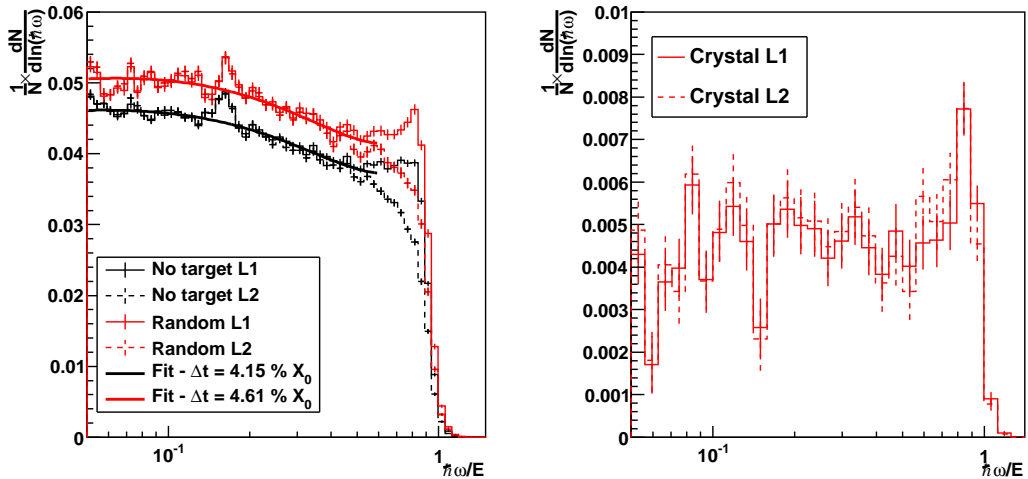}
	\caption{\label{fig:nora} (Color online) Left: Lead glass measurements for 120 GeV electrons at level 1 and 2 for no target and random position with Bethe-Heitler fits. The fits include a pile-up correction \cite{Baie99a}. Right: No target subtracted from random at level 1 and 2 to see the radiation from the non-aligned crystal.}
\end{figure*}

The main goal of this experiment was to measure a pile-up free photon spectrum from electrons subjected to the strong field in a germanium crystal and furthermore to investigate the integral radiation enhancement for electron energies corresponding to associated strong-field parameters from $\chi_s = 0.048$ to $\chi_s = 0.72$. With these parameters we probe the onset of quantum effects, where the radiation intensity in units of the classical intensity is ranging from $I_e/I_\text{cl} \simeq 0.8$ to 0.3. This has been calculated with \eref{eq:IeIcl} and the fit shown in \fref{fig:enheta}.  
In the following a thorough exposition of the data will be given together with a discussion of the observations.


\subsection{Cuts}
In the analysis we have used different cuts to clean the data. At level 0 the normalisation trigger condition is fulfilled, i.e.\ hits in Sc1 and Sc2 and no hit in ScH. A pair trigger is also used which furthermore requires a hit in either of the scintillators placed in front of the lead glass calorimeters. Since DAQ deadtime is significant with the present setup, normalisation events are scaled by 2 to get more pair events in the final data sample. This slightly biases the LG measurements but since the conversion probability is low this should only be a $\sim10\%$ effect. 

At level 1 we demand a signal in DC1 and DC2 which is used to determine the entrance angle. For the pair spectrometer measurements we furthermore require that the calibration algorithm has run successfully. To ensure that the particles are within the Baier angle of the crystal ($\theta = V_0 / m = \SI{215}{\mu rad}$) we restrict the entrance angles to be less than $\SI{100}{\mu rad}$. This is done at level 2, which is the last cut used in the LG analysis. For the pair spectrometer more cuts are used. We can use the number of particles in DC5 and DC6, the signal seen in the SSD detector, and the reconstructed vertex position in the MDX magnet. These will be further described later.


\subsection{Background radiation}
When the crystal is in a non-aligned position, the radiation spectrum should correspond to radiation from an amorphous target which is given by the Bethe-Heitler formula. The radiation spectrum from a measurement without the crystal (no target) and with the crystal in random position is shown in \fref{fig:nora}. There is a clear excess of radiation with the crystal; however from the Bethe-Heitler fits (with pile-up effects included with the corrections of Baier and Katkov\cite{Baie99a}) one sees that the excess only corresponds to $0.46\% X_0$ which is much less than the expected $0.87 \% X_0$. This discrepancy cannot be explained by an error in the crystal thickness. Since we are only hitting the central part of the crystal the thickness variation should not be more than $\pm\SI{15}{\mu m}$ around $\SI{200}{\mu m}$. This means more than $0.8 \% X_0$ at all positions. A possible explanation for the discrepancy could be a large amount of material between the MBPL magnet and the Cu converter. 
This would increase the probability of a photon converting to a pair which reduces the amount of radiation that is detected. We have made a Monte Carlo simulation to investigate this possibility, and find that it seems like a plausible explanation. The unknown material could very well be related with a disconnected helium bag. If this is the case, one would have approximately $3\% X_0$ more material from the MBPL to the converter. This would also lead to  more MCS between DC3 and DC5. From data with direct beams (MBPL turned off), which was originally used for lead glass calibrations, we observe an angular spread from DC3 to DC5 of $\sim\SI{160}{\mu rad}$, where we have fitted a gaussian distribution to the particle direction change. If we include the detector resolution of 0.4 mm this corresponds to MCS of $\sim\SI{150}{\mu rad}$ which is more than the expected MCS. These circumstances are definitely a challenge for this experiment, but the shapes of the differential radiation spectra are not significantly affected by this, since the pair conversion probability is almost constant at these energies. The extra material from the MBPL magnet to the Cu converter only decreases the amount of radiation by a certain fraction. Therefore one can correct for these effects by using the measured random radiation by the lead glass calorimeter which is $0.46\% X_0$. The enhancement spectra are not affected, since both the random and the aligned are affected by the same fraction.

For the lead glass data shown in \fref{fig:nora}, the difference between the L1 and L2 data is mainly the decrease at high photon energies. Since the difference between the random and no target spectra is almost identical for L1 and L2 data, as can be seen in \fref{fig:nora}, the change is most likely connected with background radiation from upstream the setup. The peak around $0.2E_0$ is probably also caused by radiation from upstream the setup. Similar peaks are also observed at other electron energies (see LG data in \fref{fig:ax}).

Due to beamtime limitations, only one background measurement was made at an electron energy of 120 GeV. We assume that the ratio between the random orientation spectrum and the no target spectrum is energy independent and the same for all electron energies.
For the level 2 120 GeV lead glass measurements shown in \fref{fig:nora}, we find $\d N_\text{Background}/\d \hbar\omega = 1/1.125\cdot\d N_\text{random}/\d \hbar\omega$. For the pair spectrometer we have found a similar value for this ratio.


\subsection{Simulations of the pair spectrometer}
\begin{figure*}[tb]
\centering
\includegraphics[width=0.8\textwidth]{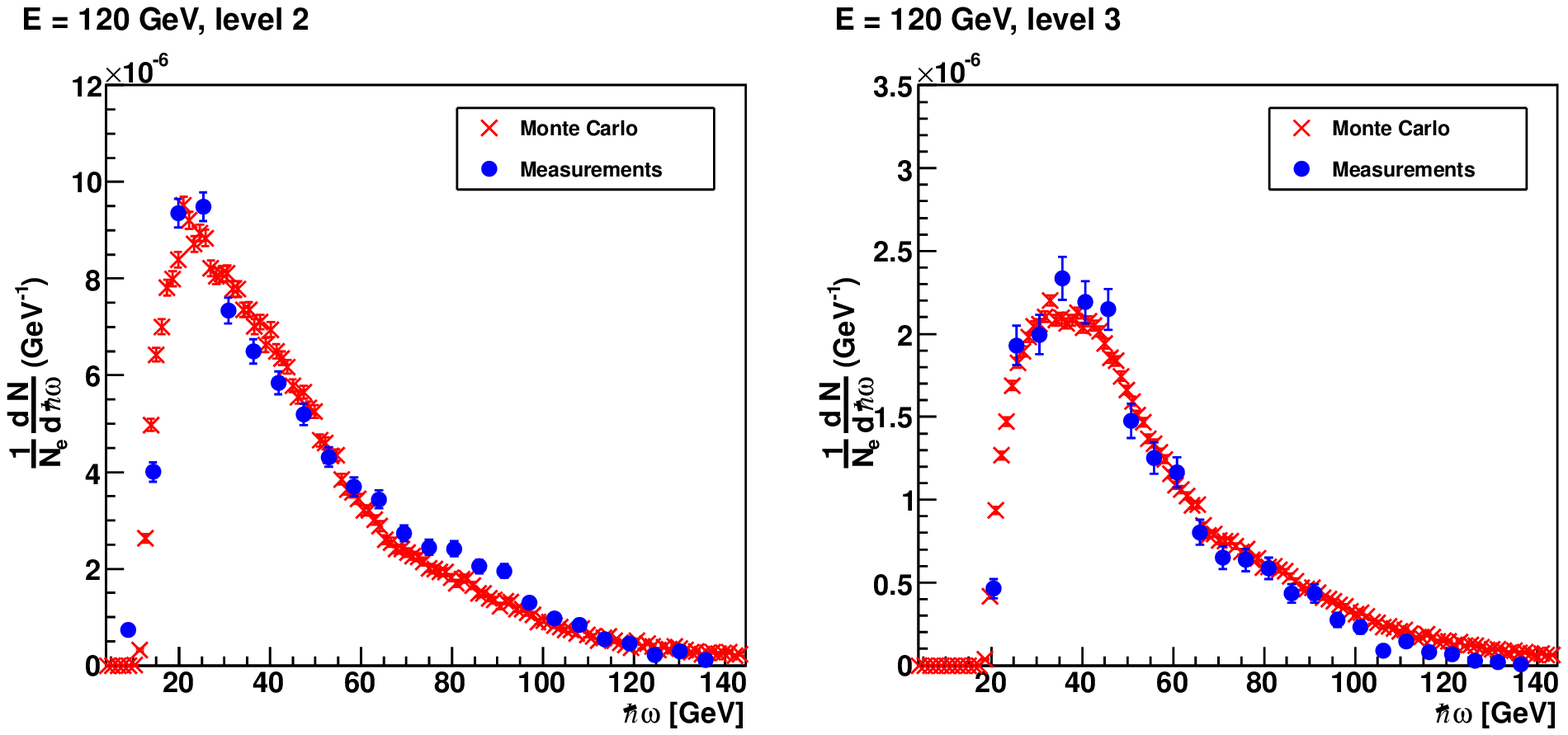}%
\caption{\label{fig:MCcomp} Monte Carlo simulations (red crosses) and random data (blue dots) for 120 GeV electrons at level 2 (left) and level 3 (right). At level 2 the simulation has been scaled with 0.11 and at level 3 with 0.045. }%
\end{figure*}

\begin{table*}[tb]
\begin{tabular}{ccc}
\hline
Level & Pre Cut & Post Cut \\
\hline
0 & Norm trigger & No cut \\
1 & Pre Cut 0 and hit in DC1 and DC2 & Pair spec. algorithm has run \\
2 & Pre Cut 1 and entrance angle less than $\SI{100}{\mu rad}$ & Post Cut 1 \\
3 & Pre Cut 2 & Post Cut 1 and 2 hits in both DC5 and DC6 \\
\hline		
\end{tabular}
\caption{\label{tab:cuts} Cuts used for the PS analysis. Pre cuts are used to limit the events used in normalisation and post cuts are used to define good events.}
\end{table*}  

The detection efficiency of the lead glass calorimeter is close to 1. This is not the case for the pair spectrometer and therefore it is crucial to include it. We have determined the efficiency by comparing the measured random spectra with a Monte Carlo simulation of the spectrometer. In the simulation, we include detector resolution, multiple coulomb scattering, and detector geometry. We let an incoming electron emit photons according to the Bethe-Heitler radiation probability. If one or more photons are emitted, we always let one and only one photon pair convert. Since the pair-conversion probability above 1 GeV is to a good approximation independent of the photon energy, one can later scale the simulation by the conversion probability. The geometry of the spectrometer results in a detection threshold of around 10 GeV depending on the cuts and MDX field used. In the simulation we assume that the crystal contributes only $0.46 \% X_0$, since this is what is measured with the lead glass calorimeter. In \fref{fig:MCcomp} we show 120 GeV data and simulations at level 2 and 3. The cuts used can be found in \tref{tab:cuts}. The simulations have been scaled by a normalisation factor. This factor is $0.114$ at L2 and $0.045$ at L3, but varies slightly for the different electron energies.  Since the factor includes the conversion probability of the Cu converter, it is expected to be close to 0.11, since the converter thickness is 14\%$X_0$. This is also the case for the L2 data. The large difference between L2 and L3 is unexplained, but the shapes of the MC and the data still agree very well. Nevertheless, we do not think this poses a serious problem and the spectrometer setup is considered understood.

\begin{figure*}[tb]
\centering
\includegraphics[width=0.8\textwidth]{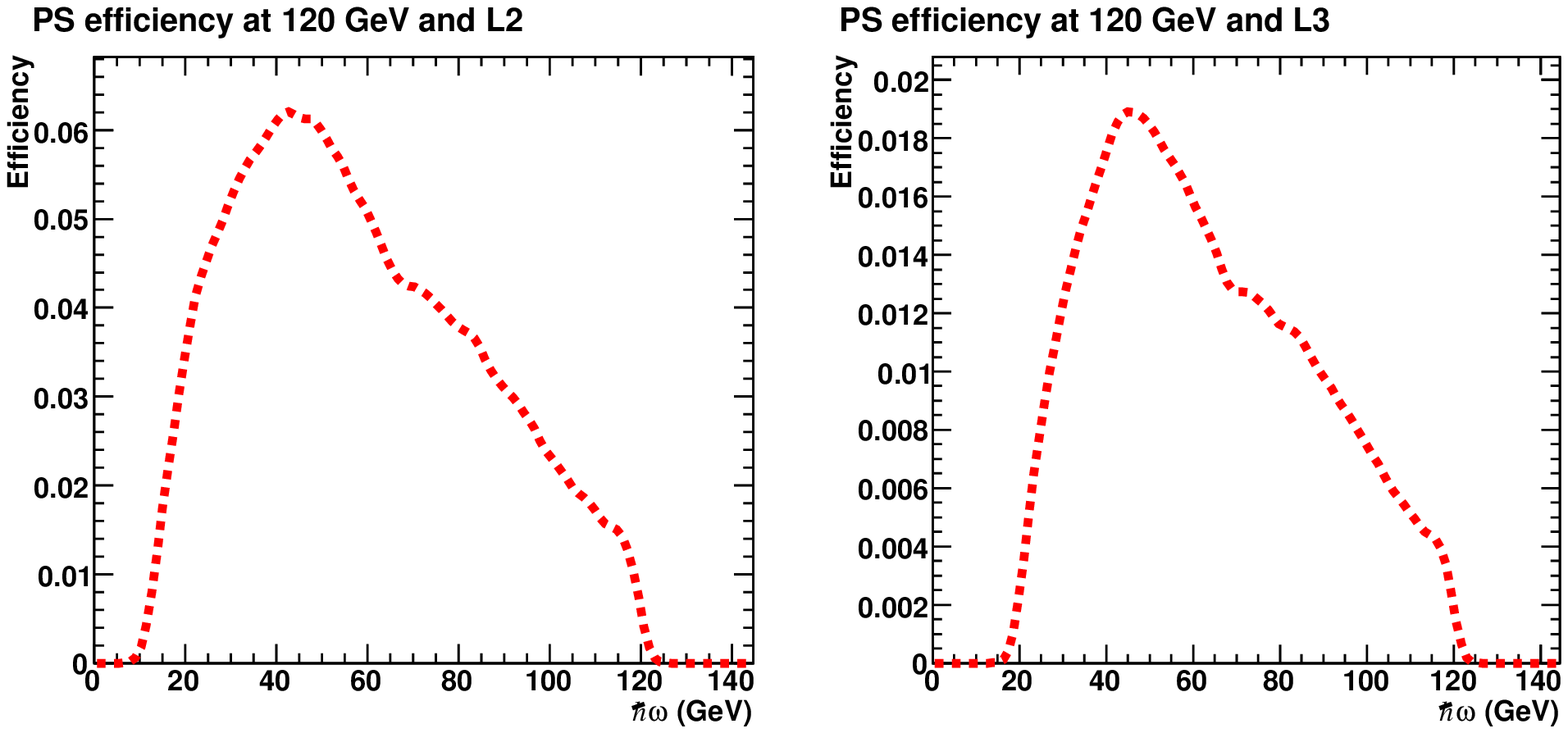}
\caption{\label{fig:PSeff} Pair spectrometer efficiency at level 2 and 3 for 120 GeV data. The efficiency includes the conversion probability from the Cu converter.}
\end{figure*}  

The level 3 cut is only used to ''clean'' the data and do not effect the normalization. The difference between L2 and L3 is the demand of two hits in both DC5 and DC6. This removes the low energy part of the data as also seen in \fref{fig:MCcomp}. Furthermore, the double hit restriction seems to improve the energy determination of the spectrometer significantly. The effect of additional cuts used to clean the PS data has been investigated. For example cuts in the SSD signal and cuts on the deflection vertex in the MDX. These however do not affect the spectral shape significantly and they have not been used in the final data analysis.

The pair spectrometer detection efficiencies can be found from the normalisation factors and the simulations and is plotted in \fref{fig:PSeff}. With the efficiency correction we can directly compare the measured PS radiation spectra to the theoretical calculations. However, the radiation enhancement is the ratio between the axial radiation and random radiation and is therefore independent of the detection efficiency. This is a big advantage since the results in that case do not rely on MC simulations.


\subsection{Differential radiation spectra}

\begin{figure*}[tb]
  \includegraphics[width=0.32\textwidth]{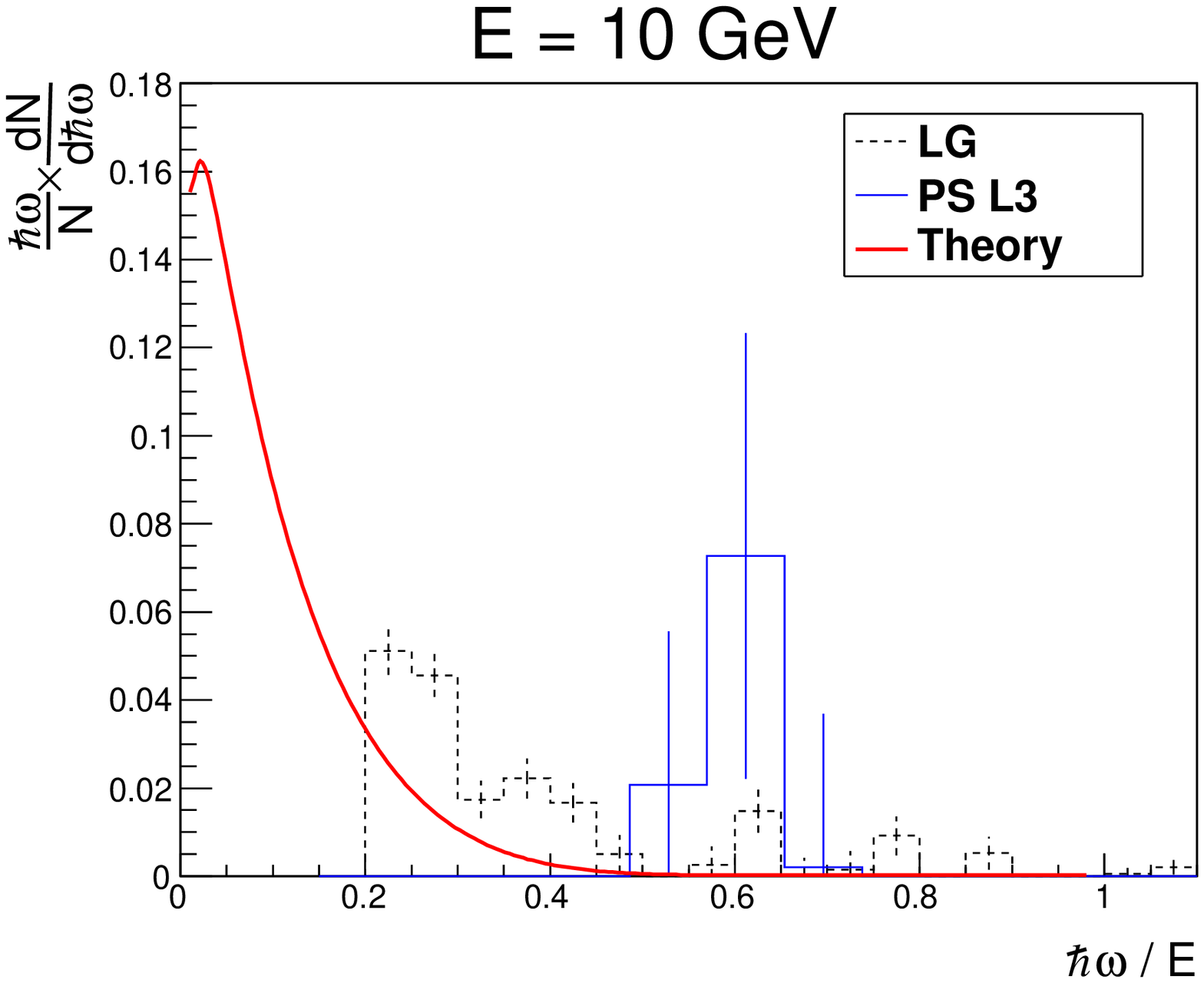}	
	\includegraphics[width=0.32\textwidth]{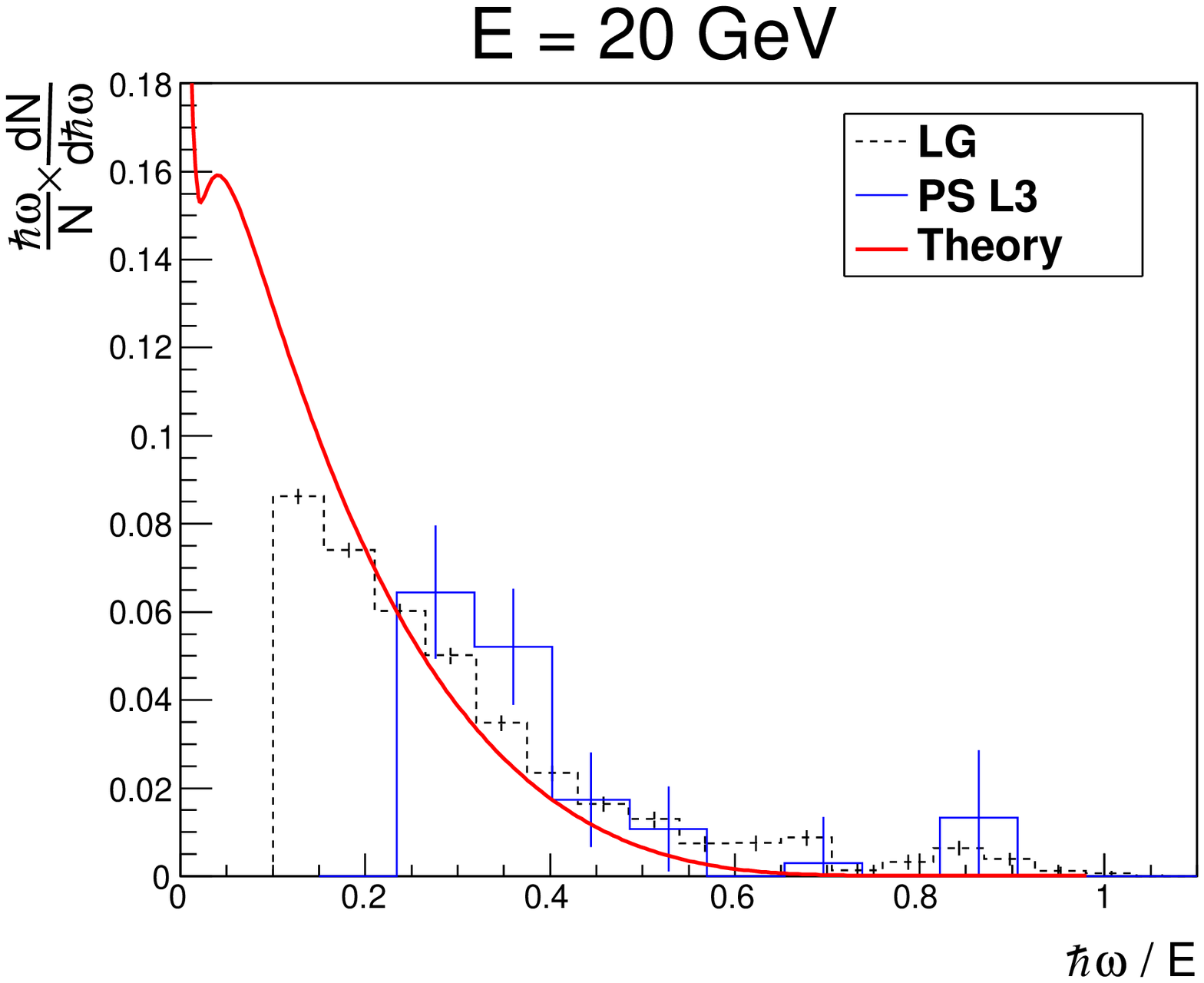}
	\includegraphics[width=0.32\textwidth]{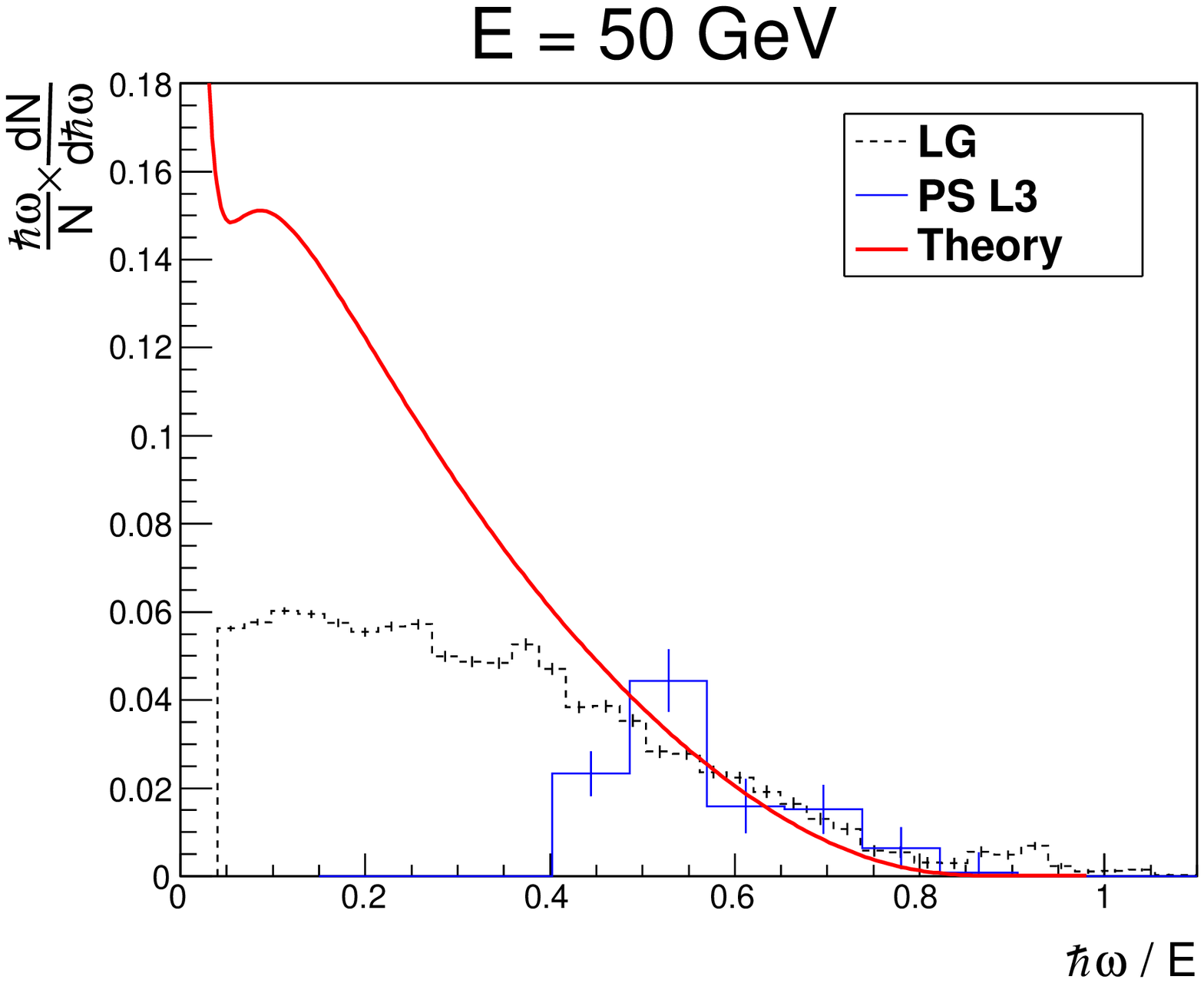}
	\includegraphics[width=0.32\textwidth]{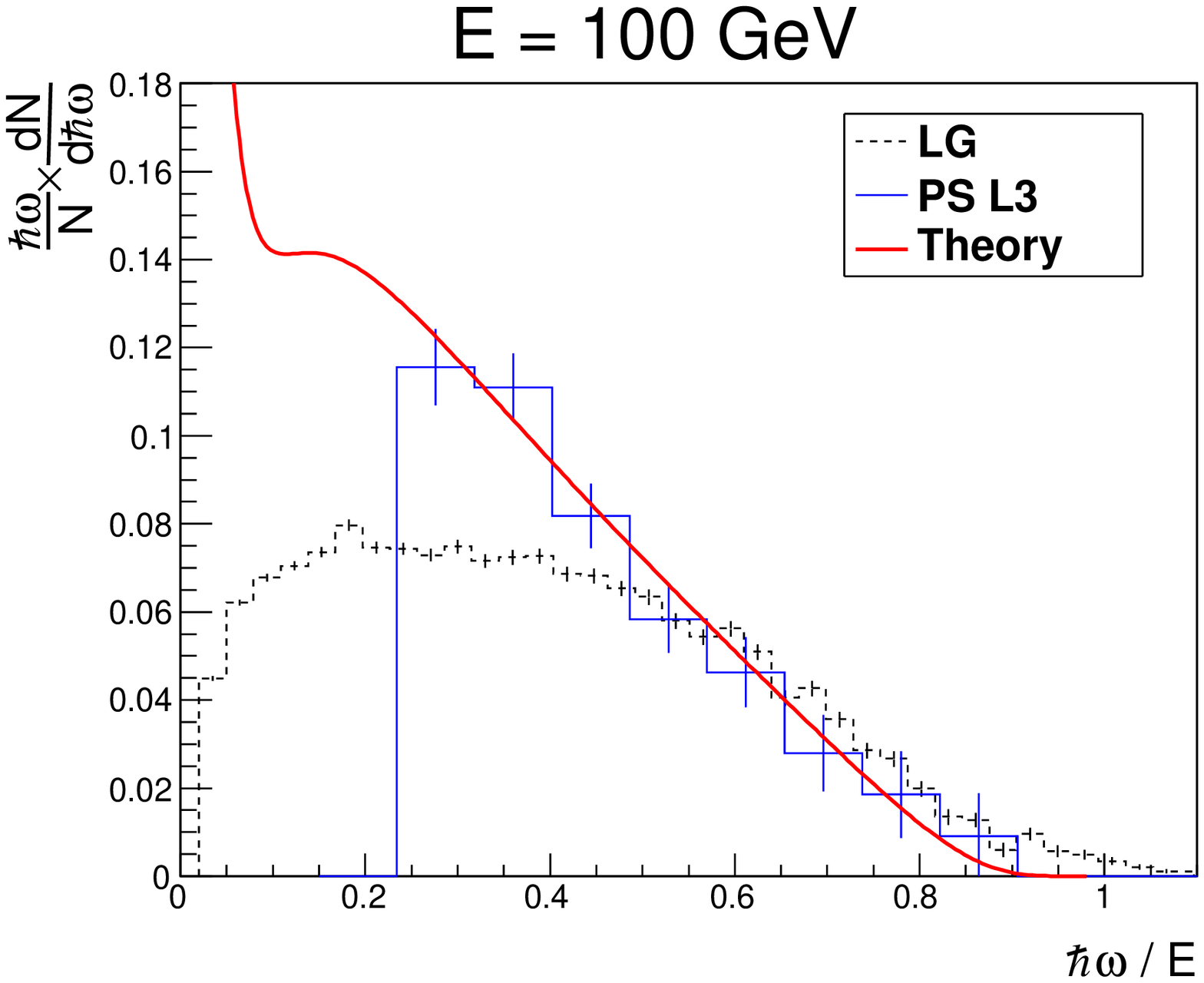}
	\includegraphics[width=0.32\textwidth]{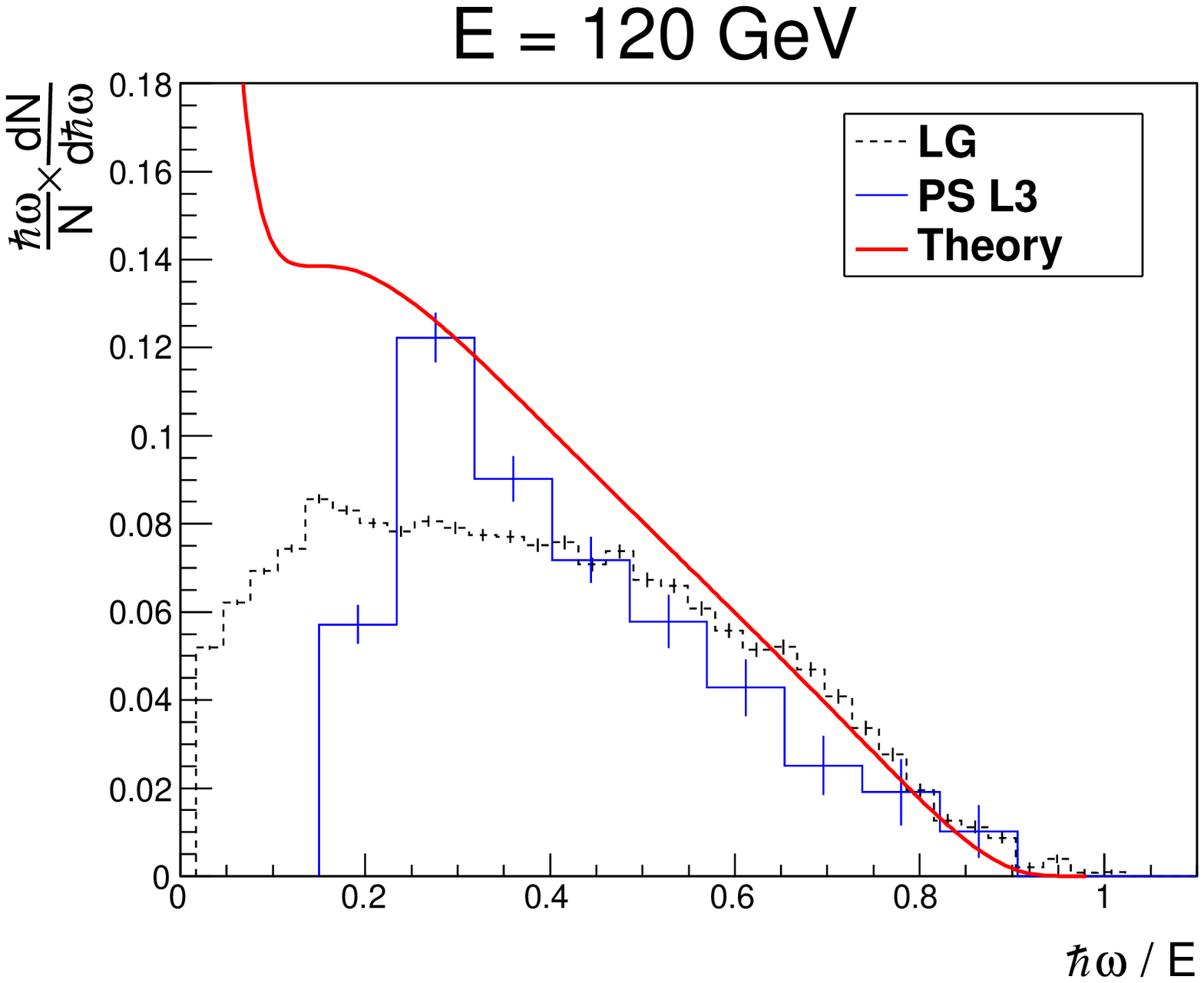}
	\includegraphics[width=0.32\textwidth]{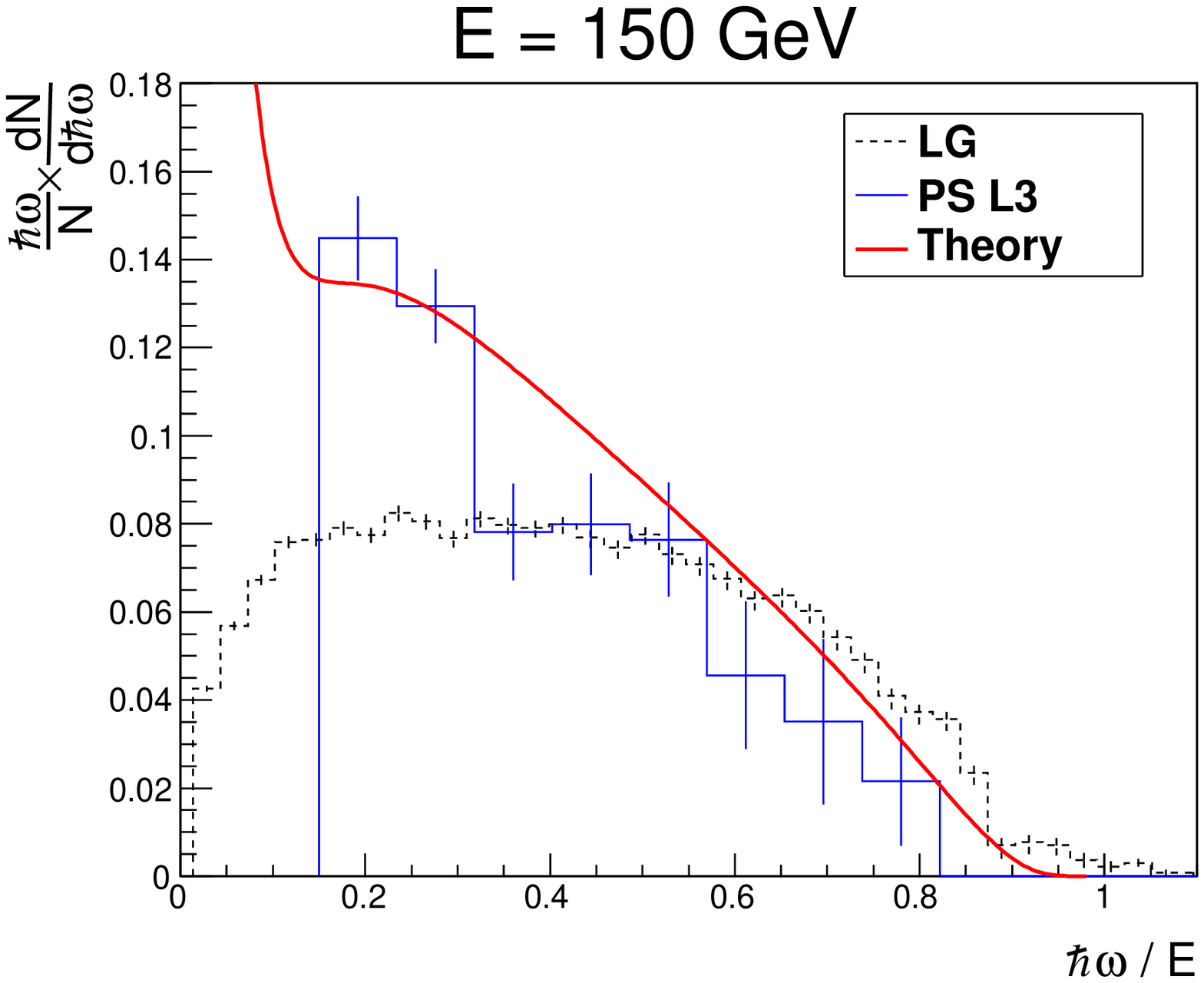}
	\caption{\label{fig:ax} Differential radiation spectra for 10, 20, 50, 100, 120 and 150 GeV electrons along the Ge $\langle 110 \rangle$ axis for LG and PS measurements. Theoretical calculations based on formulas by Baier \textit{et al.} are also plotted. The error bars only represent statistical uncertainties.}
\end{figure*}

The calculated differential radiation spectra are compared to both the LG and the PS measurements. The calculated spectra have been averaged over entrance angles from 0 to $\SI{165}{\mu rad}$. The value used for the upper limit was determined by the applicability of the theoretical formula. Since the beam divergence is $\sim\SI{100}{\mu rad}$ and the position of the axis is somewhat uncertain we assume a uniform particle distribution over 0 to $\SI{165}{\mu rad}$. The differential spectra are only slightly dependent on the entrance angle at these values and the assumption of a uniform distribution does not affect the theoretical results significantly. \fref{fig:enhteo} also shows that the radiation intensity is only slightly affected by the electron entrance angle.
The axial PS L3 data are plotted in \fref{fig:ax}. The shape of the measured spectra generally agree with the theoretical calculations. The data are shown with only statistical errors. 

The lead glass measurements are consistently below the PS data and the theoretical curve at low photon energies. We expect this to be caused by pile-up which is also consistent with the resulting increase observed at higher energies. For 10 GeV the PS statistics is very poor. The 120 GeV data are consistently below the theoretical values. This is almost certainly caused by a slightly misaligned crystal as mentioned earlier. 

\subsection{Enhancement}
\begin{figure}[tb]
	\includegraphics[width=\columnwidth]{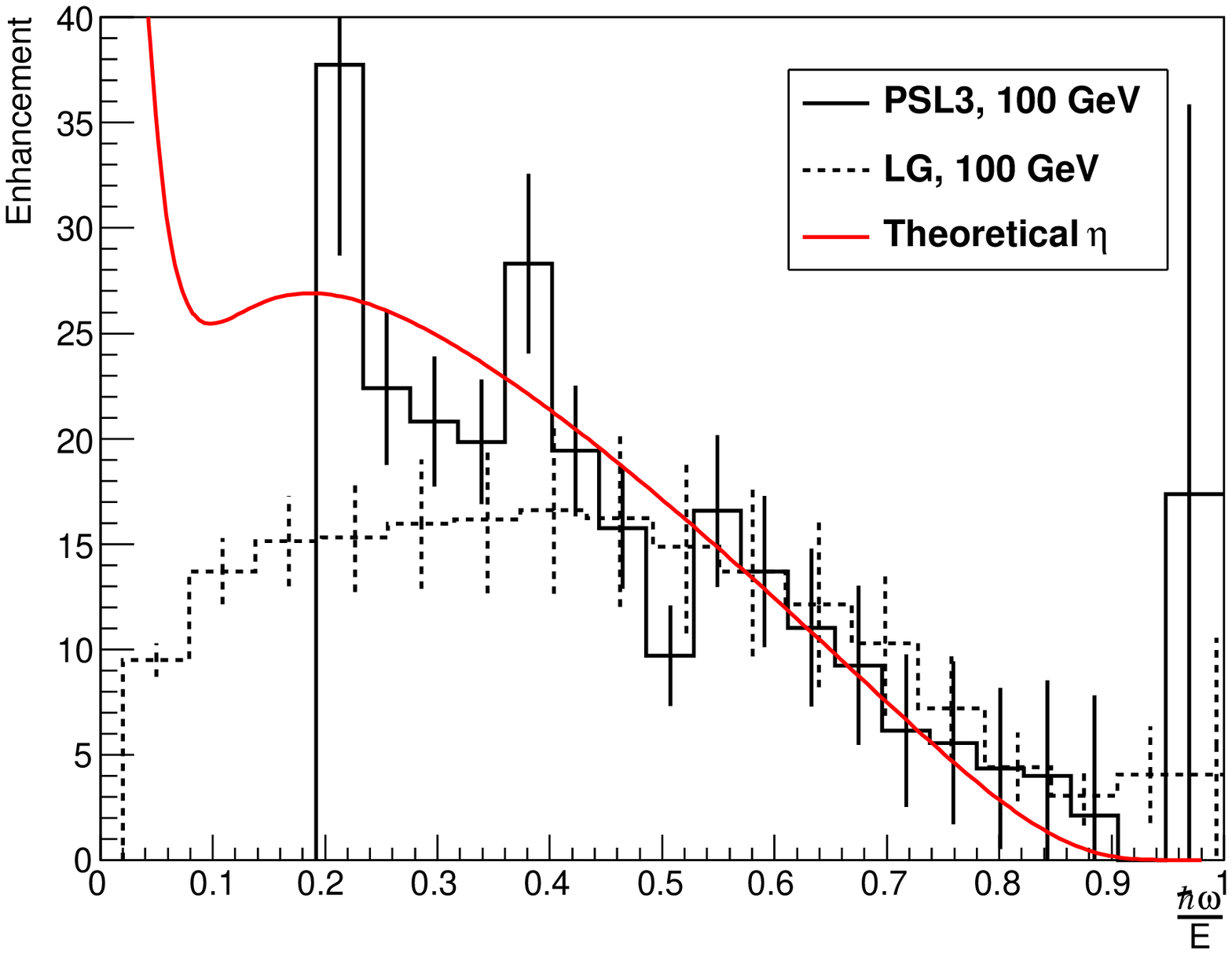}
	\caption{\label{fig:enh100} Radiation enhancement as function of photon energy for PS L3 (black solid) and LG data (black dashed). The theoretical enhancement (red line) calculated with Baiers formulas agrees well with the PS data. }
\end{figure}

The differential radiation enhancement is defined as
\begin{equation}
\eta(\hbar\omega) = \frac{\d N_\text{axial}/\d \hbar\omega}{\d N_\text{random}/\d \hbar\omega}
\label{eq:eta}
\end{equation}
where both spectra have had the background subtracted. The differential enhancement for 100 GeV PS L3 data is shown in \fref{fig:enh100} together with the LG data and a theoretical curve. The curves are similar to \fref{fig:ax} since the random spectrum is close to constant.

\begin{figure}[tb]
	\includegraphics[width=\columnwidth]{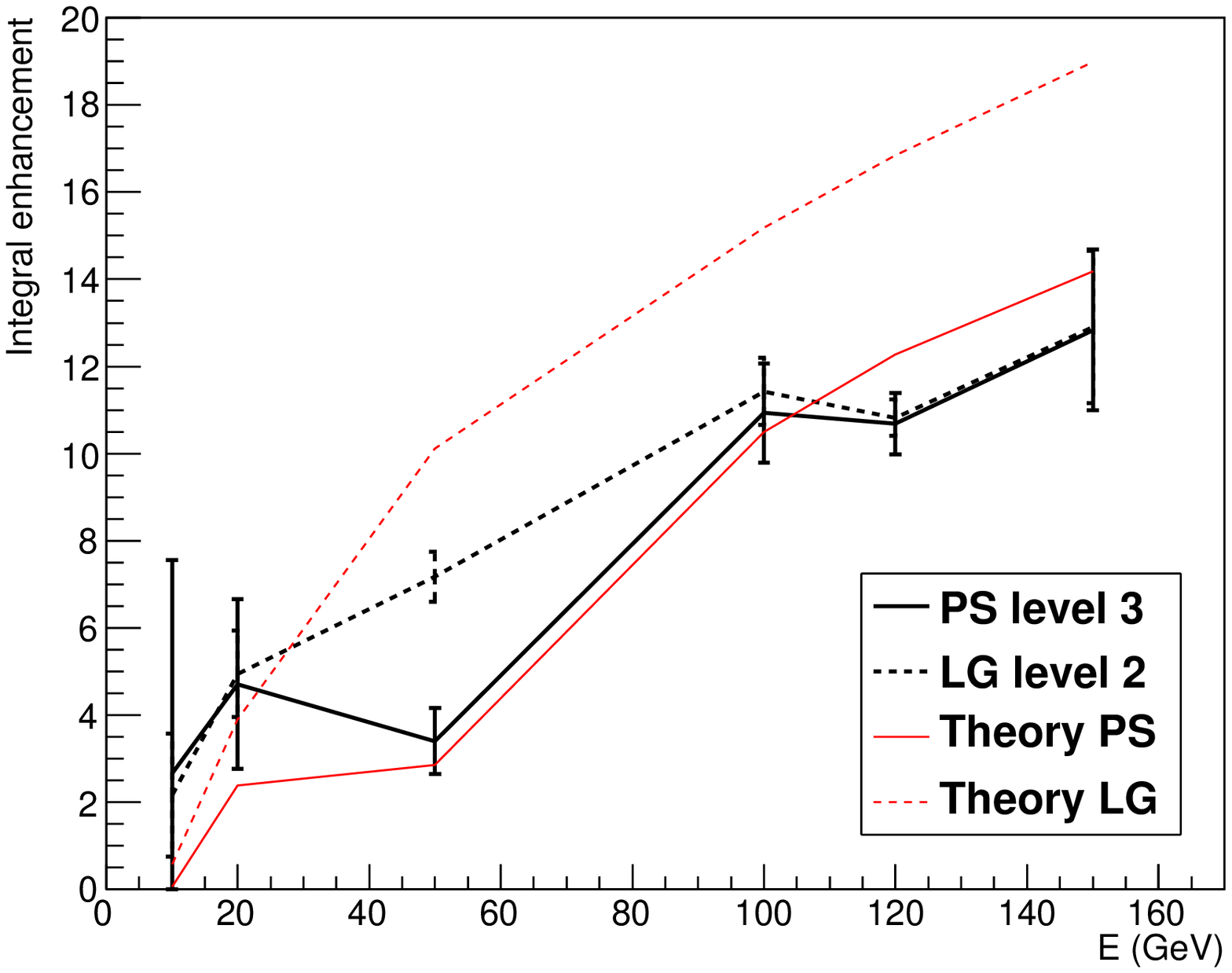}
	\caption{\label{fig:enhall} The integral radiation enhancement as a function of electron energy for PS L3 (black solid) and LG data (black dashed). The theoretical integral enhancement calculated with Baiers formulas and including detection energy thresholds for PS L3 (red solid) and LG (red dashed) are also shown.}
\end{figure}

The integral enhancement is plotted in \fref{fig:enhall} for both LG and PS data. Theoretical calculations of the enhancements that have been corrected for the detection energy thresholds, are also plotted. These enhancements are lower than those plotted in \fref{fig:enhteo} since the threshold is 2 GeV for LG measurements and 18 GeV for PS data at level 3 (3 GeV for the 10 and 20 GeV data where the MDX was run at lower current).
A good agreement between the PS data and theory can be seen, but there are significant discrepancies for the LG data. The drop in enhancement from 100 GeV to 50 GeV is connected to the detection thresholds. The discrepancies between theory and the LG data is probably caused by pile-up. A $\chi^2$ analysis of the PS L3 data gives $\chi^2/\text{ndf} = 7.99/6$ and a $\chi^2$ probability of 24\%.

\begin{figure}[tb]
		\includegraphics[width=\columnwidth]{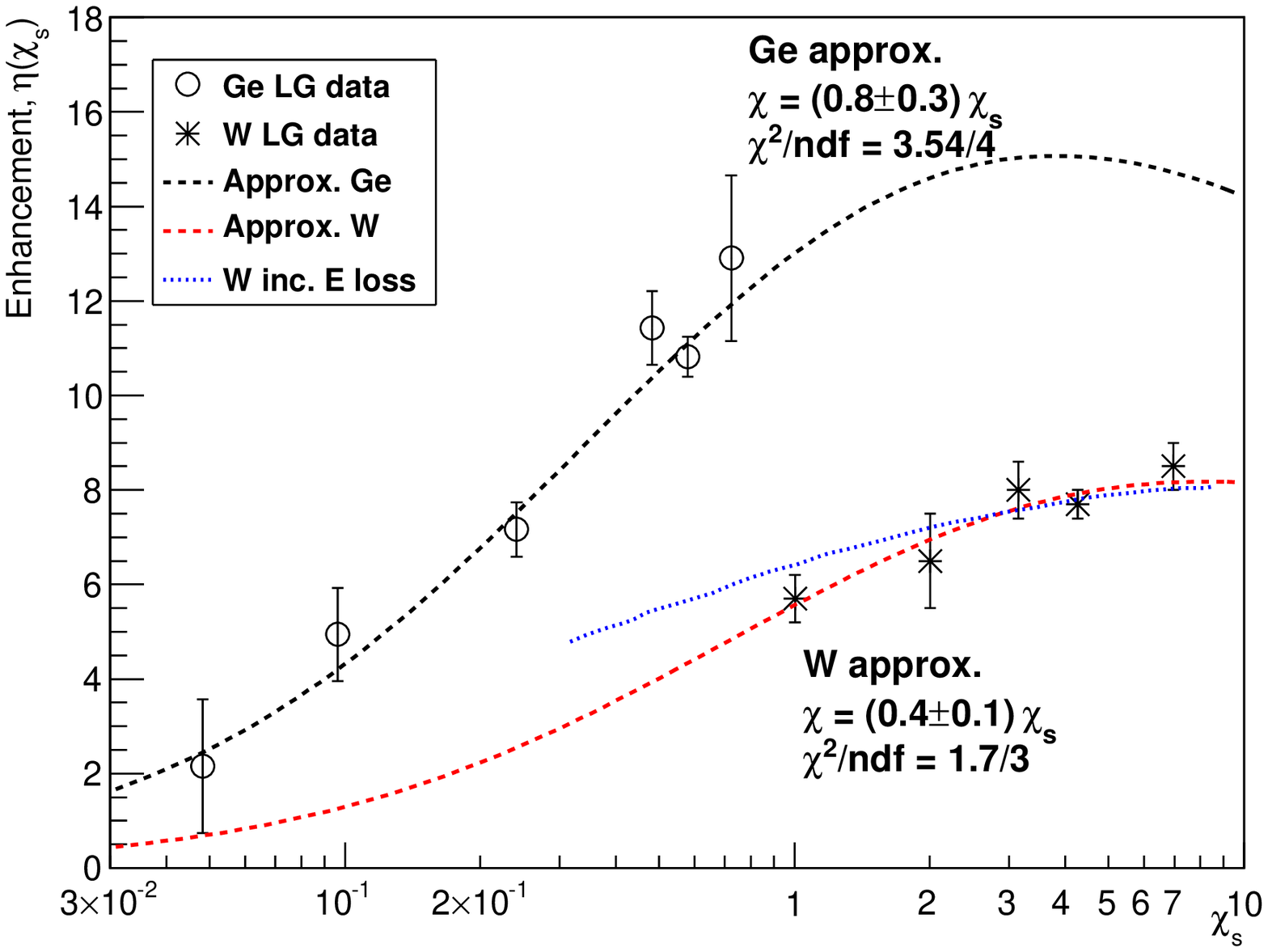}
	\caption{ 	\label{fig:enheta} The radiation enhancement as a function of the associated strong-field parameter $\chi_s$ for Ge$\langle 110 \rangle$ crystal detected by a lead glass calorimeter and W$\langle 111\rangle$ lead glass measurements \cite{Kirs01a}. The data have been fitted by the approximate quantum suppression formula, \eref{eq:IeIcl}, as described in the text. A theoretical calculation including energy loss for the W data is also plotted.}
\end{figure}

In this experiment we probe associated strong-field parameters of $\chi_s$ that are below unity and hence the onset of quantum effects. In a previous experiment by Kirsebom \textit{et al.}\cite{Kirs01a} values of the associated strong-field parameters above unity was probed in a tungsten crystal. These results are plotted together with our lead glass measurements in \fref{fig:enheta}. One cannot directly compare the two data sets, since the enhancement not only depends on $\chi_s$ but also on the crystal structure. 

We wish to model the $\chi_s$ dependence of the enhancement with expressions calculated for single-field radiation. This is not strictly correct but may give an approximate relation between $\chi$, $\chi_s$, and the quantum suppression. Baier \textit{et al.}\cite{Baie98} have found approximate expressions for the quantum suppression of synchroton radiation as a function of $\chi$.
\begin{equation}
\frac{I_e}{I_\text{cl}} = (1+4.8(1+\chi)\ln(1+1.7\chi)+2.44\chi^2)^{-2/3}.
\label{eq:IeIcl}
\end{equation} 
Since $I_\text{cl}\propto \chi^2$ and $I_\text{BH}\propto E \propto \chi$ the radiation enhancement is given by $\eta(\chi) = I_e/I_\text{BH} \propto \chi I_e$. \eref{eq:chi} shows that the local $\chi$ is proportional to $\chi_s$. We have therefore fitted $\eta(A\chi_s)=B \chi_s I_e(A\chi_s)$ where $A$ and $B$ are fitting parameters. The results are shown in \fref{fig:enheta}.

Baier \textit{et al.} have also made a calculation that includes energy loss during the passage of the W crystal (blue line) \cite{Baie06} which is plotted. Such calculations have not been made for the Ge crystal.

\begin{figure}[tb]
		\includegraphics[width=\columnwidth]{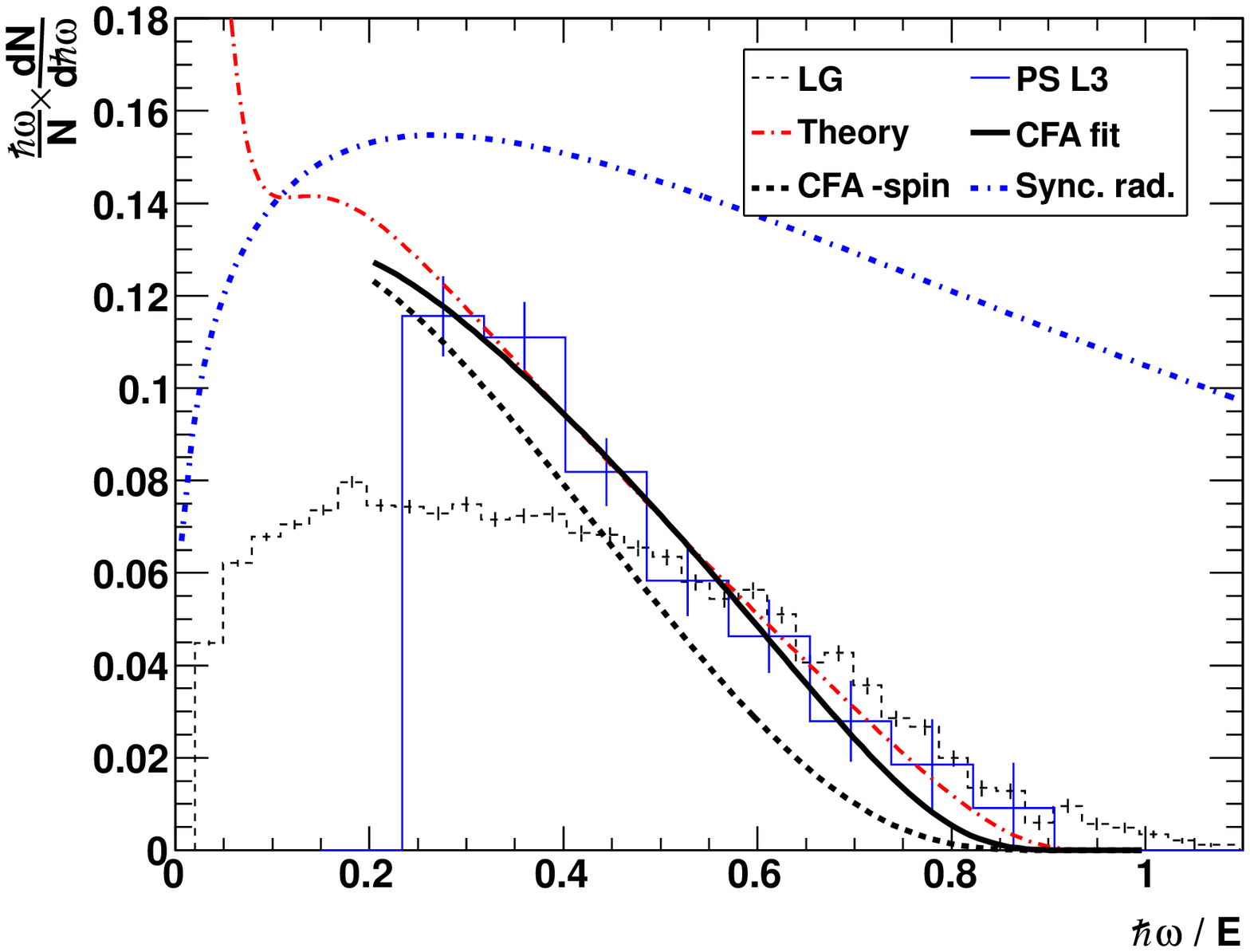}
\caption{\label{fig:SRcomp} 100 GeV PS data with theoretical calculations. The CFA has been fitted to the data (black line) and we find $\chi = 0.68\pm0.16$. The CFA without the spin contribution has been plotted with the parameters found from the CFA fit (black dotted) and the classical synchroton radiation spectrum has been plotted for the same field (blue).}
	
\end{figure}

As mentioned at the beginning of the article one of the central objectives of this experiment was to observe the change in radiation spectrum when quantum recoil and spin-flip transitions affect the process. In \fref{fig:SRcomp} we compare the 100 GeV PS data to several theoretical calculations. Besides the curve from \fref{fig:ax}, we have fitted a CFA calculation to the data. The fit parameters are the strong-field parameter $\chi$ and a scaling factor. We find $\chi = 0.68\pm0.16$. With this field we plot the CFA theory without the spin-flip contribution and the classical radiation spectrum. Spin-flip transitions are only a small contribution at this field but the difference between the classical spectrum and the measurements is drastic and the classical formula is clearly inadequate at these fields.


\section{Conclusion}
This is the first experiment to probe the associated strong-field parameter interval from $\chi_s = 0.048$ to $\chi_s = 0.723$ and measure pile-up free photon spectra from channeled electrons. We have compared theoretical calculations based on the constant field approximation to measurements of the differential radiation spectra from an aligned germanium crystal. With the electron energies used in this experiment we investigate the onset of quantum suppression of synchroton radiation. This is relevant for possible future electron-positron colliders. The present experimental results are in agreement with theories.

Finally, we note that our previous experiment on 'tridents' \cite{Esbe10}, where a factor 2-3 disagreement with theory was found, was performed with a setup very similar to that presented here, and with the same type of target crystal. The present findings support our interpretation that the explanation for the disagreement reported there is likely to be due to effects not included in the theory.


\section{Acknowledgments}

We are very grateful for the strong support from P.B. Christensen and P. Aggerholm(DPA, Aarhus). Financial support from the Danish Natural Science Research Council (FNU/NICE) is acknowledged.


\end{document}